\RequirePackage{lineno}
\documentclass[aps,prd,superscriptaddress,twocolumn, showpacs, showkeys, amsmath, amssymb, floatfix]{revtex4} 
\usepackage[colorlinks=true, citecolor=green, filecolor=blue, linkcolor=blue, urlcolor=blue]{hyperref}
\usepackage{color}
\usepackage{graphicx}
\usepackage[perpage]{footmisc}
\usepackage{amssymb, amsmath}
\usepackage{subfigure}
\usepackage{graphicx}
\usepackage{capt-of}

\usepackage{rotating}

\begin{document}

\title{First Searches for Axions and Axion-Like Particles with the LUX Experiment}


\author{D.S.~Akerib} 
\affiliation{Case Western Reserve University, Department of Physics, 10900 Euclid Ave, Cleveland, OH 44106, USA}
\affiliation{SLAC National Accelerator Laboratory, 2575 Sand Hill Road, Menlo Park, CA 94205, USA}
\affiliation{Kavli Institute for Particle Astrophysics and Cosmology, Stanford University, 452 Lomita Mall, Stanford, CA 94309, USA}

\author{S.~Alsum} 
\affiliation{University of Wisconsin-Madison, Department of Physics, 1150 University Ave., Madison, WI 53706, USA}

\author{C.~Aquino} 
\affiliation{SUPA, School of Physics and Astronomy, University of Edinburgh, Edinburgh EH9 3FD, United Kingdom}

\author{H.M.~Ara\'{u}jo} 
\affiliation{Imperial College London, High Energy Physics, Blackett Laboratory, London SW7 2BZ, United Kingdom}

\author{X.~Bai} 
\affiliation{South Dakota School of Mines and Technology, 501 East St Joseph St., Rapid City, SD 57701, USA}

\author{A.J.~Bailey} 
\affiliation{Imperial College London, High Energy Physics, Blackett Laboratory, London SW7 2BZ, United Kingdom}

\author{J.~Balajthy} 
\affiliation{University of Maryland, Department of Physics, College Park, MD 20742, USA}


\author{P.~Beltrame} 
\affiliation{SUPA, School of Physics and Astronomy, University of Edinburgh, Edinburgh EH9 3FD, United Kingdom}

\author{E.P.~Bernard} 
\affiliation{University of California Berkeley, Department of Physics, Berkeley, CA 94720, USA}
\affiliation{Yale University, Department of Physics, 217 Prospect St., New Haven, CT 06511, USA}

\author{A.~Bernstein} 
\affiliation{Lawrence Livermore National Laboratory, 7000 East Ave., Livermore, CA 94551, USA}

% ...


\begin{abstract}

\noindent
The first searches for axions and axion-like particles with the Large Underground Xenon experiment are presented. Under the assumption of an axio-electric interaction in xenon, the coupling constant between axions and electrons $g_{Ae}$ is tested using data collected in 2013 with an exposure totalling 95 live days~$\times$118~kg. A double-sided, profile likelihood ratio statistic test excludes $g_{Ae}$ larger than 3.5$\times$10$^{-12}$ (90\% C.L.) for solar axions. Assuming the Dine-Fischler-Srednicki-Zhitnitsky theoretical description, the upper limit in coupling corresponds to an upper limit on axion mass of 0.12 eV/$c^{2}$, while for the Kim-Shifman-Vainshtein-Zhakharov description masses above 36.6 eV/$c^{2}$ are excluded. For galactic axion-like particles, values of $g_{Ae}$ larger than 4.2$\times$10$^{-13}$ are excluded for particle masses in the range 1--16~keV/$c^{2}$. 
These are the most stringent constraints to date for these interactions.

\end{abstract}

\pacs{12.60.-i, 14.80.Va, 95.35.+d, 95.30.Cq}
\keywords{Axions, Dark Matter, LUX}

\maketitle

\section{Introduction}

\noindent
The standard model of particle physics has long been thought to be incomplete as it is, for example, unable to 
explain dark matter, the observed matter-antimatter asymmetry of the universe, or the hierarchy problem. 
Another major weakness is the lack of a natural mechanism to explain the 
absence of charge-parity (\textit{CP}) violation in strong interactions.  
A solution, introduced by Peccei and Quinn~\cite{jr:PQ}, postulates an  additional 
global symmetry $U(1)_{PQ}$ that is spontaneously broken at some large energy scale, $f_{a}$. This 
generates a Nambu-Goldstone boson, the Weinberg-Wilczek {\it axion}~\cite{jr:WeinbergAxion, jr:WilczekAxion}, with a field that
transforms 
as $a(x) \rightarrow a(x) + \alpha f_{a}$,
where $\alpha$ is the phase of the introduced scalar field. 
If there is more than one global symmetry and, therefore, more than one Nambu-Goldstone boson, 
the particle corresponding to the excitation of the field combination is then the axion.  Axions arising from symmetry breaking at 
electroweak scales have been discounted, having been ruled out by experimental searches~\cite{jr:KimCarosi}, but axions that result from a much higher energy scale, so-called ``invisible'' axions~\cite{jr:DFSZ, jr:KSVZ1, jr:KSVZ2}, remain viable. 
In addition to QCD axions, particle excitations of the fields orthogonal to this field combination are called Axion-Like-Particles (ALPs),
and indeed, numerous string-theory driven models predict ALP candidates~\cite{jr:Witten84, jr:Conlon06, jr:string, jr:Cicoli12}.

Both axions and ALPs 
make 
 interesting dark matter candidates~\cite{jr:Abbott83}: they are nearly collisionless, neutral, nonbaryonic, and may be present 
 in sufficient quantities to provide the expected dark matter density. Axions may have been produced as a nonthermal relic by the misalignment mechanism~\cite{jr:InvAxion, jr:Dine83} and while very light, are predicted to be produced essentially at rest, thus satisfying the criteria for cold dark matter.
There are also possible thermal production mechanisms~\cite{jr:ThermalAxion}, although these are unlikely to result in significant contributions to the dark matter.  ALPs may have been present during the early phases of the Universe, produced as stable or long-lived particles that are now slowly moving within our Galaxy~\cite{jr:steffen2009}. 

Production of axions may arise in stellar environments
leading to a constant rate of emission from stars. From the Sun, this provides a second possible source of axion signal, 
but the consistency of stellar behaviour with models that exclude axion emission also leads to tight constraints on their existence~\cite{jr:WhiteDwarfs, jr:RedGiants, jr:SolarNu}. Additional constraints arise from searches for axion couplings to photons via the Primakoff effect
~\cite{jr:ADMX, jr:CAST}. 
Axions and ALPs are also expected to couple with electrons, so can be probed with a wider range of experimental techniques, such as instruments with germanium and xenon active targets~\cite{jr:arisaka2013, jr:XENON100}. Here we present searches for axio-electric coupling with the LUX experiment
for two specific scenarios: i) QCD axions emitted from the Sun, and ii) keV-scale galactic ALPs that could constitute the gravitationally bound dark matter.

\section{Signal expectation in LUX}

The Large Underground Xenon experiment (LUX) provides sensitivity to dark matter in the form of weakly interacting massive particles (WIMPs), reporting, for example, the most sensitive limits to date for spin-independent and spin-dependent WIMP-neutron interactions for masses above 4~GeV/$c^{2}$~\cite{jr:firstLUX,jr:reanaLUX,jr:SD_Run04_LUX,jr:combinedLUX}. LUX is a dual-phase xenon time-projection chamber (TPC) consisting of a low-radioactivity titanium vessel partially filled with liquid xenon such that above the liquid a layer of gaseous xenon is maintained. A vertical electric field of 181~V/cm is established via a gate grid placed within the gas layer, 
and a cathode at the base of the liquid. The detector has an active target mass of 250~kg.  
Energy deposited by incident radiation creates a primary scintillation signal, called $S1$, and ionization charge. The latter, when drifted vertically in an electric field to produce an electroluminescence signal in the gas phase, leads to a delayed signal, called $S2$. Both signals are detected by photomultiplier tubes (PMTs), 61 viewing the TPC from above and 61 from below. The location at which an energy deposition occurred may be reconstructed from the distribution of signal sizes in the PMTs, which gives the position in the horizontal plane. The standard deviations of the reconstructed coordinates have a statistical contribution of 10~mm at the S2 threshold
due to Poisson fluctuations in the numbers of detected photons. To this, a 5~mm systematic contribution is added, as estimated from events that
arise from the well-defined wall position~\cite{jr:reanaLUX}.
The period of delay (0-324~$\mu$s) between the $S1$ and the $S2$ then gives the vertical position, with a resolution of 0.9~mm~\cite{jr:reanaLUX}. 
The ionization threshold is sufficiently low to allow observation of single electrons emitted from the liquid surface, giving a very low energy threshold for experimental searches. 
A detailed description of the detector and its deployment at the Sanford Underground Research Facility may be found in Ref.~\cite{jr:LUXdet2013}.  

Importantly, axion or ALP interactions in LUX 
would result in additional events within the electron-recoil class of events, identified principally by the ratio of $S2$ to $S1$ signal size.  This is in contrast 
to searches for WIMPs that are conducted within the nuclear recoil band. Moreover, whereas the nuclear recoil band is essentially background free
(dominated in fact by leakage from the electron recoil band), the electron recoil band is populated significantly, 
with contributions from gamma rays and beta particles from radioactive contaminations within the xenon, from the detector instrumentation, and from external environmental sources. 
Data presented here, and their analysis, come from the period April 24th to September 1st, 2013, 
with a total exposure consisting of 118~kg fiducial mass over a 95 live days period.

Axion and ALP searches rely on the so-called {\it axio-electric effect}~\cite{jr:AxioelectricEffect0, jr:AxioelectricEffect, jr:AxioelectricEffect2} 
\begin{equation}
\sigma_{Ae} = \sigma_{pe} (E_{A}) \frac{g_{Ae}^{2}}{\beta_{A}} \frac{3 E_{A}^{2}}{16 \pi \alpha_{em} m_{e}^{2}} \left( 1 - \frac{\beta^{2/3}_{A}}{3} \right),  
\label{eq:axeXS}
\end{equation}
where $\sigma_{pe}$ is the photoelectric cross section on the target material (xenon), $g_{Ae}$ is the coupling constant between axion or ALP and electron, 
$\alpha_{em}$ is the fine-structure constant, $m_{e}$ is the mass of the electron, and 
$\beta_{A}$ and $E_{A}$ are the velocity and the energy of the axion.

\begin{figure}[h!]
\includegraphics[width=0.5\textwidth]{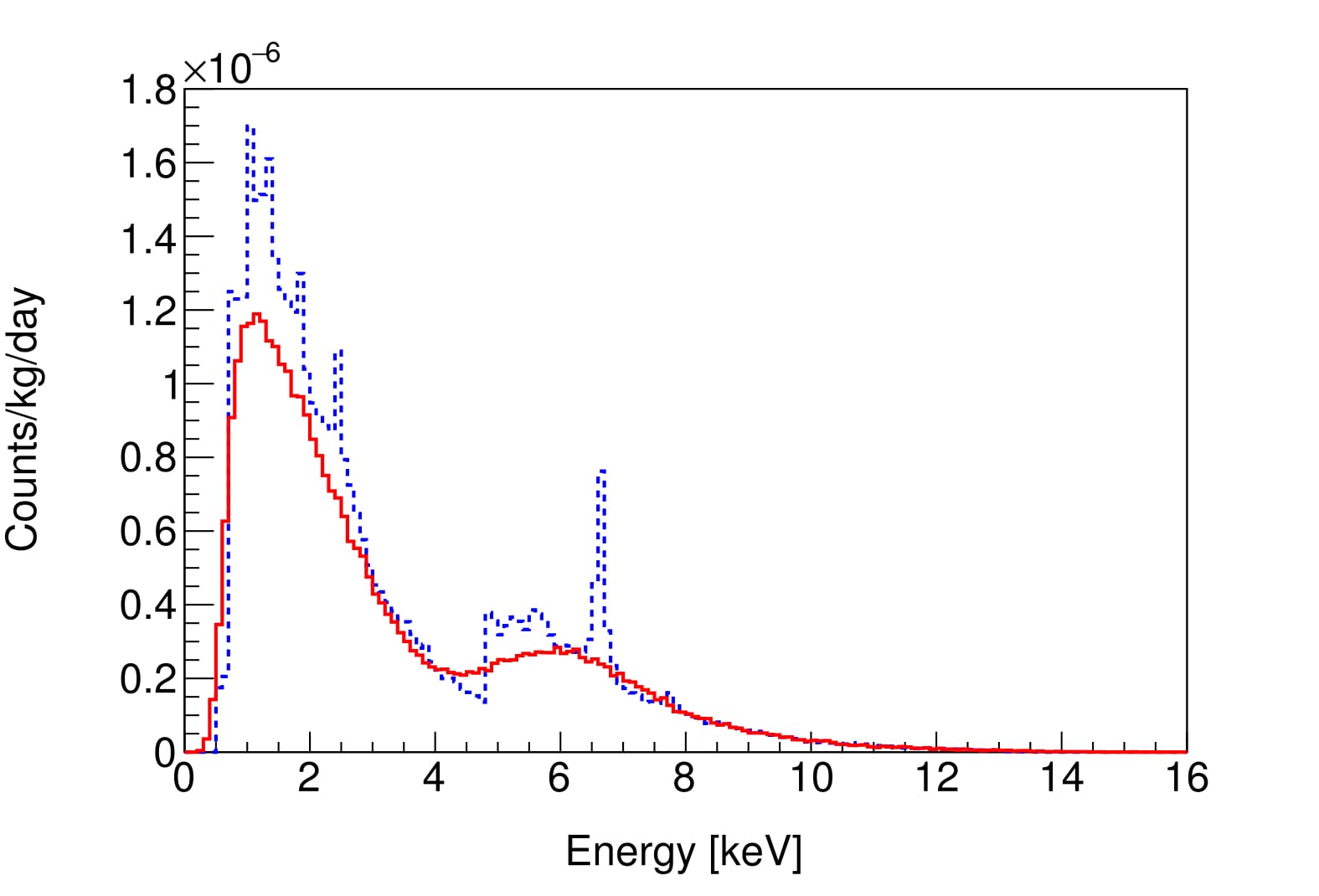}
\caption{Dashed blue distribution: expected energy spectrum from a massless solar axion, assuming a coupling $g_{Ae}=10^{-12}$. The shape arises from the combination of a continuous contribution to the axion flux due to bremsstrahlung and Compton scattering, together with features associated with atomic recombination and deexcitation effects. Solid red distribution: the expected LUX experimental solar axion energy spectrum, as modelled with NEST~\cite{jr:NEST1, jr:NEST2, jr:yields}.}
 \label{fig:spectrumSolar}
\end{figure}

Two signal sources are considered here: axions produced and emitted from the Sun, and primordial ALPs within the Galaxy. In the first case, 
Redondo~\cite{jr:redondo2013} has estimated the solar axion spectral shape, assuming massless axions.
The flux is dominated by contributions from atomic recombination and deexcitation (that introduce features associated with atomic shell structure), bremsstrahlung and Compton scattering (both of which contribute smoothly), and is presented as the dashed blue line
in Fig.~\ref{fig:spectrumSolar}, for an arbitrary choice of axion coupling constant. 
The flux, as estimated for zero axion mass, is still valid without heavy corrections for masses smaller than 1 keV/$c^{2}$ since the total energy is
dominated by kinetic energy.
The solar axion is therefore approximated to be massless, but note that these models cover theoretically interesting phase space, including the region for which axions provide a solution to the strong CP problem. 
Such a signal detected in LUX would be modified by detector resolution and efficiency effects~\cite{jr:LUXeffER}. These have been modelled 
with the Noble Element Simulation Technique (NEST) package~\cite{jr:NEST1, jr:NEST2, jr:yields} 
with the resulting 
expected solar axion energy spectrum presented as the solid red distribution in Fig.~\ref{fig:spectrumSolar}. 

%

In the case of ALP interactions within a detector, because the ALPs are expected to be essentially at rest within the galaxy, 
axio-electric absorption leads to electron recoils with kinetic energy equal to the mass of the ALP.  Interactions of this type therefore produce a monoenergetic spectral feature.

\section{Background model}

The detector design, its location deep underground, and its construction from radiopure materials contribute to ensuring a low 
rate of events from background radioactivity. Moreover, xenon attenuates radiation relatively strongly ($Z$=54, density $\sim$ 3~g/cm${^2}$) which, 
combined with the ability to accurately reconstruct the position of the interaction point, allows fiducialization away from local sources of background 
such as the walls that surround the xenon target, the PMTs and the cathode. 

\begin{figure*}[ht!]
\centering
\begin{minipage}[l]{1.0\textwidth}
    \includegraphics[width=0.4\textwidth]{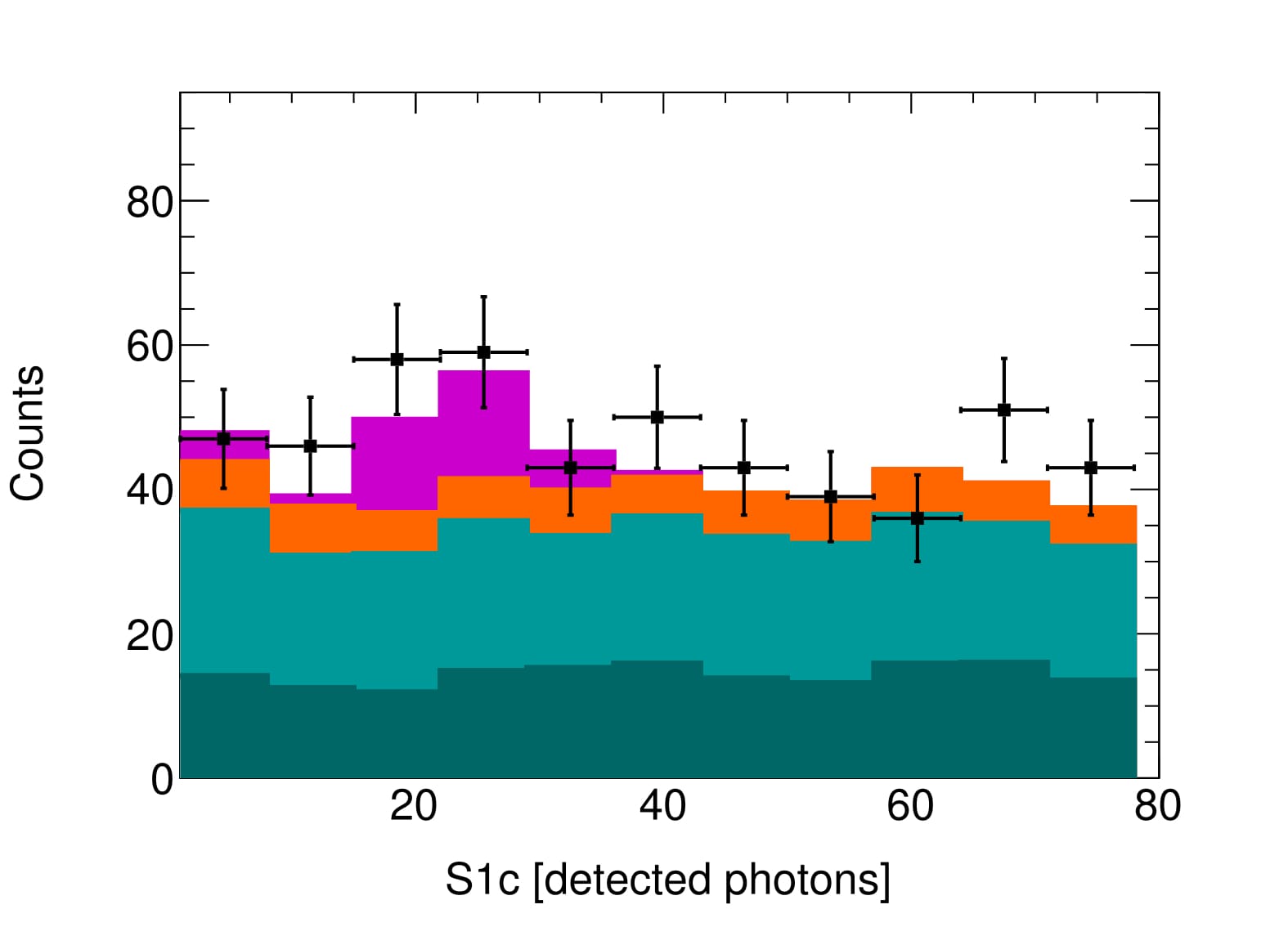}
    \includegraphics[width=0.4\textwidth]{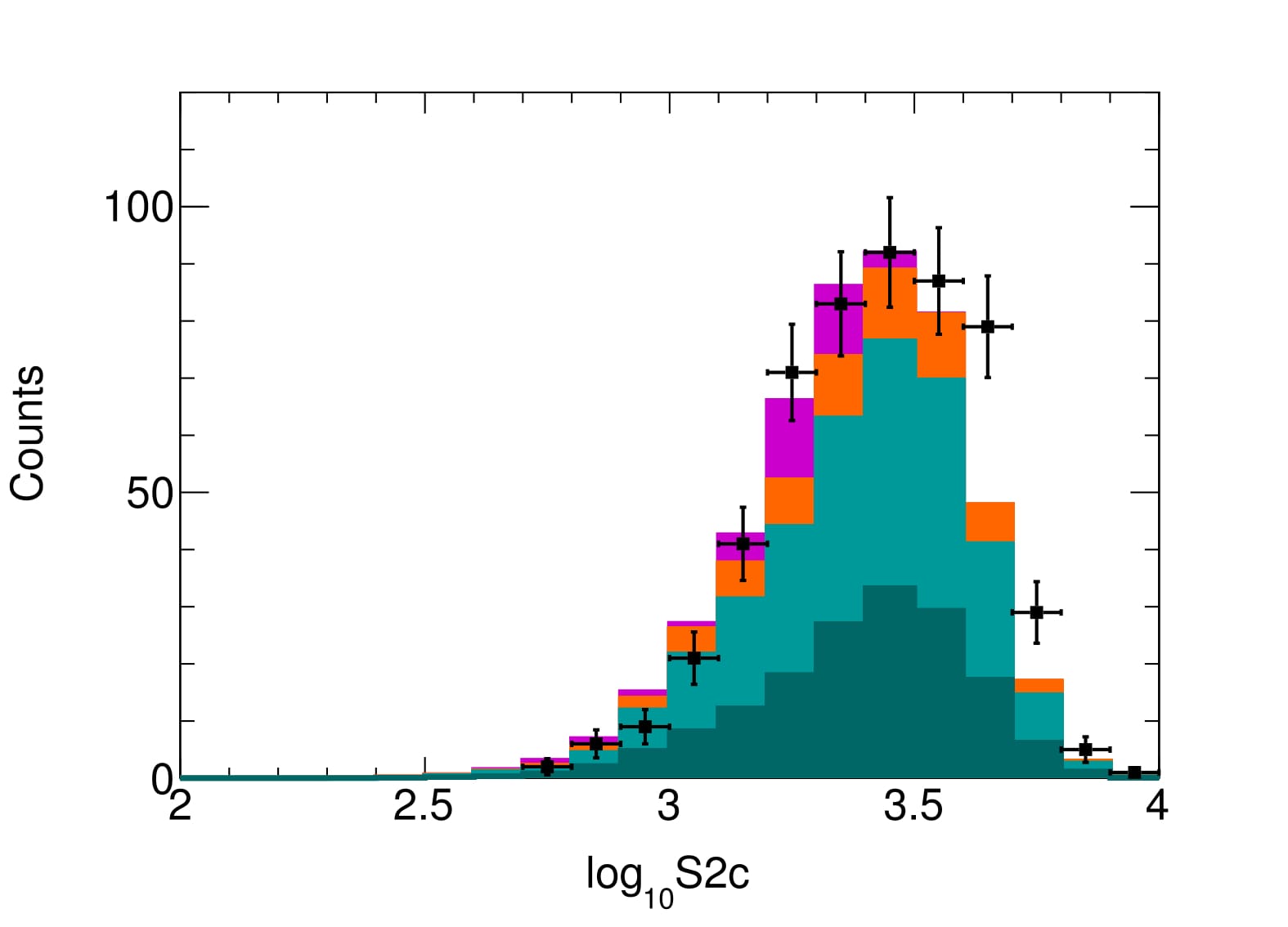}
\end{minipage}
\begin{minipage}[l]{1.0\textwidth}
    \includegraphics[width=0.4\textwidth]{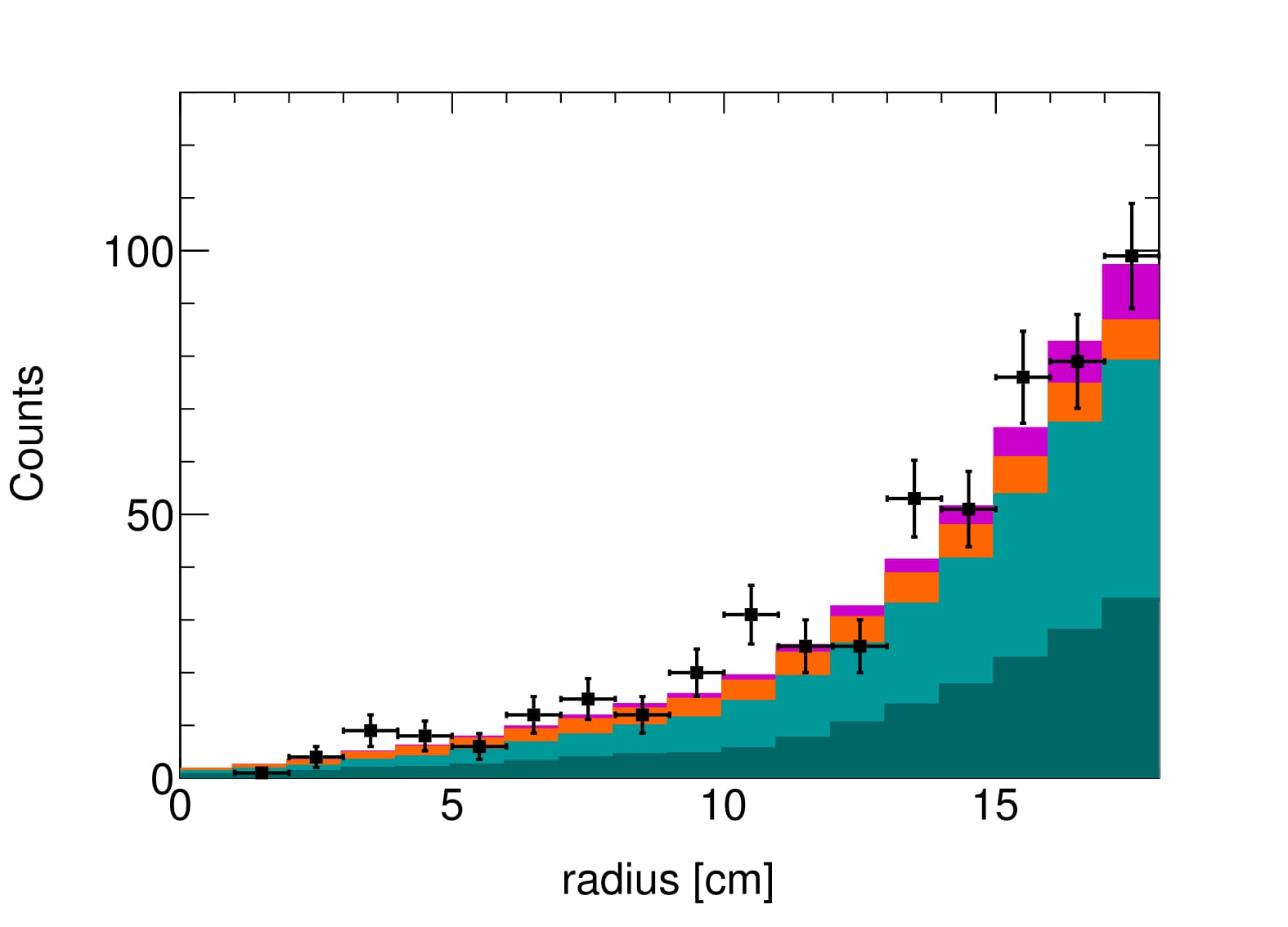}
    \includegraphics[width=0.4\textwidth]{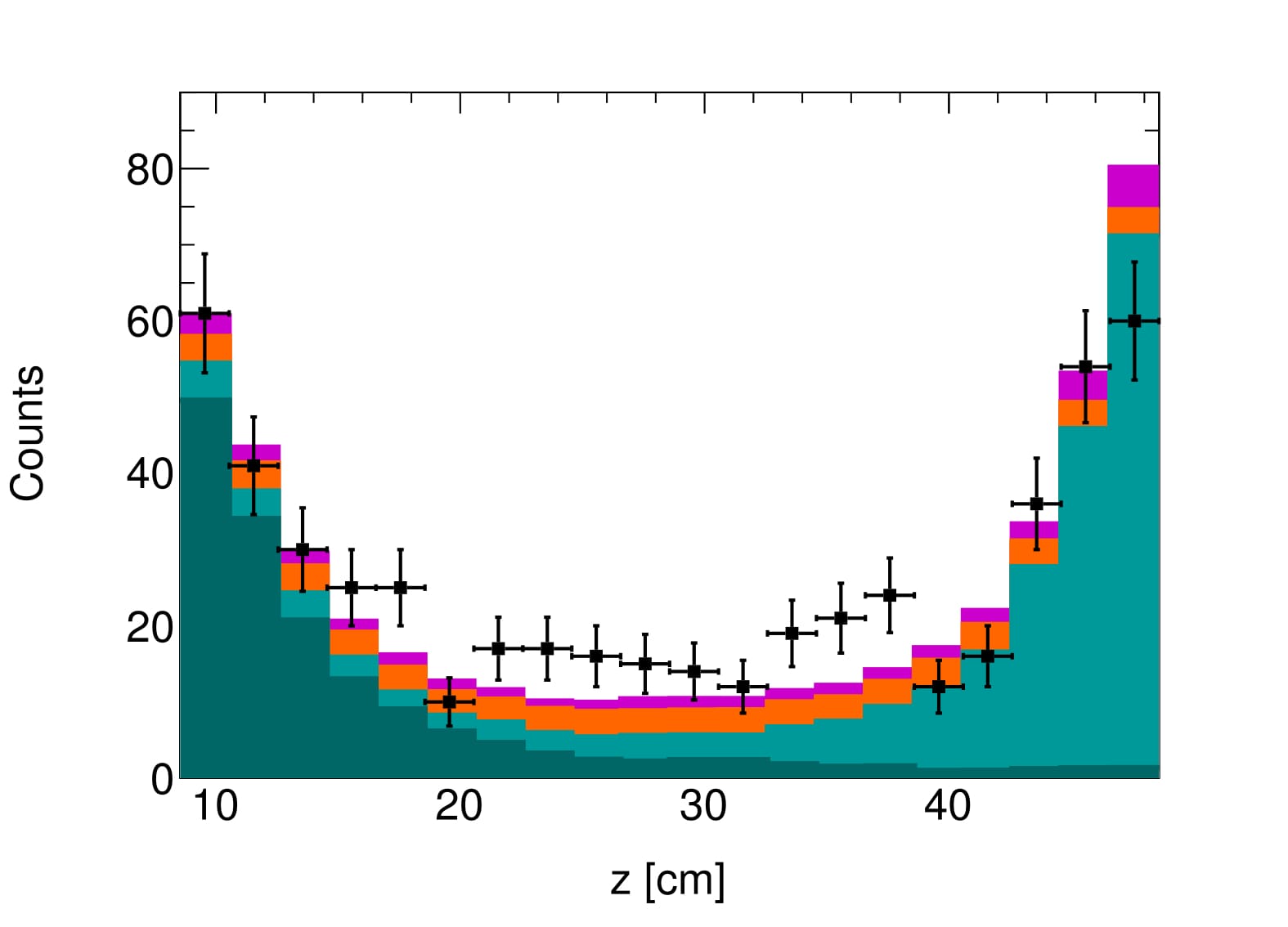}
\end{minipage}
\caption{LUX 2013 electron recoil data (filled black squares, with error bars) together with the background model, comprised of contributions from low-$z$-origin $\gamma$ rays (dark green), other $\gamma$ rays (light green), $^{85}$Kr or Rn-daughter contaminants in the liquid xenon undergoing $\beta$ decay (orange) and x rays due to $^{127}$Xe (purple).   The four panels show the distributions in terms of the four parameters used in the
analysis: $S1_{c}$ (top left),  log$_{10}S2_{c}$ (top right),  radial coordinate (bottom left) and vertical coordinate (bottom right). The number of counts in each background component is based on independent assay results and measurements, with no additional scaling.}
\label{fig:BG_obs}
\end{figure*}

Figure~\ref{fig:BG_obs} presents, for the fiducial volume and the energy region of interest, the LUX 2013 data, together with the background model.
Radiogenic backgrounds are estimated as in Ref.~\cite{jr:LUXbg} 
and lead to a contribution from Compton scattering of $\gamma$ rays from detector component radioactivity (light green). An additional 
$\gamma$-ray contribution arising from heavily down-scattered emission from $^{238}$U chain, $^{232}$Th chain, and $^{60}$Co decays in the center of a large copper block below the PMTs is also included~\cite{jr:reanaLUX} (dark green).
Further significant contributions arise from $^{85}$Kr and Rn-daughter contaminants in the liquid xenon undergoing $\beta$ decay with no accompanying $\gamma$ rays detected (orange), and x rays emitted following those $^{127}$Xe electron-capture decays where the coincident $\gamma$ ray escapes the xenon (purple). Each background contribution has been estimated from modelling measured impurity levels, and no scaling 
has been performed.
The four observables used in the subsequent statistical analysis are modelled: the prompt scintillation ($S1$), the base 10 logarithm of the proportional ($S2$) signal, the radius ($r$), and depth ($z$) of the event location. 
$S1$ pulses are required to have two-PMTs in coincidence and an $S1$ value in the range 1--80 detected photons; the $S2$ signal is required to be in the range 100--10000 detected photons.
A radial fiducial cut is placed at 18~cm and the range in $z$ is set to be 48.6--8.5~cm above the faces of the bottom PMTs. The resulting fiducial volume has been calculated as in Ref.~\cite{jr:firstLUX}.

\begin{figure}[ht!]
    \includegraphics[width=0.5\textwidth]{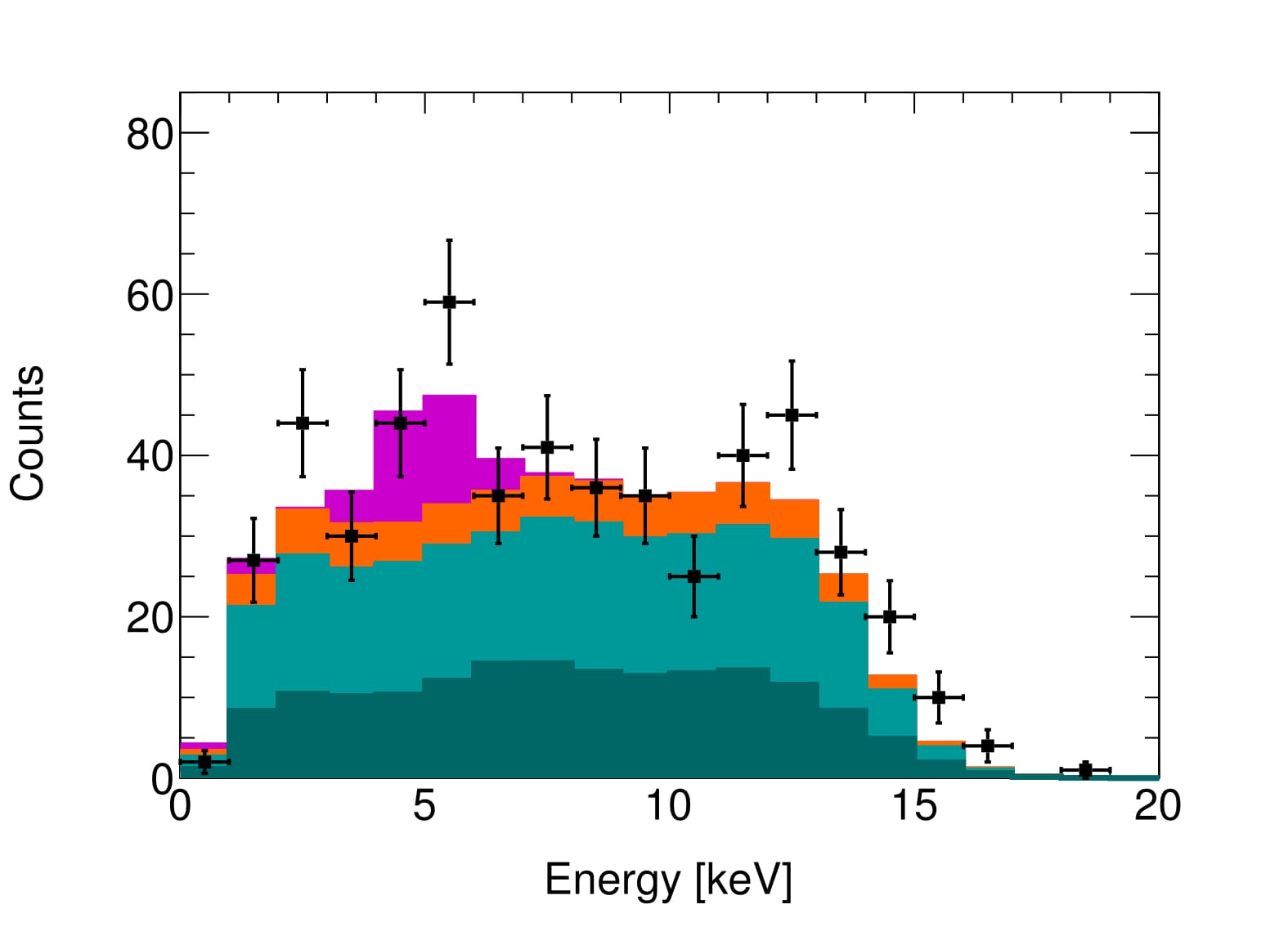}
    \caption{Energy spectrum of the LUX 2013 electron recoil background. Data are filled black squares with error bars; the individual contributions to the background model are the stacked colored histograms: low-$z$-origin $\gamma$ rays (dark green), other $\gamma$ rays (light green), $^{85}$Kr or Rn-daughter contaminants in the liquid xenon undergoing $\beta$ decay (orange), x rays due to $^{127}$Xe (purple). The number of counts in each background component is based on independent assay results and measurements, with no additional scaling. The cutoff at higher energies is due to the requirement on $S1$ signal size.}
    \label{fig:BG_energy}
\end{figure}	

Figure~\ref{fig:BG_energy} shows the background model and LUX 2013 data as a function of recoil energy, with energy reconstructed as $ E = [S1_{c}/g_{1} + S2_{c}/(\epsilon g_{2})] W $. Here, $S1_{c}$ is the $S1$ signal size corrected to equalize the response throughout the active volume to the response at the center of the detector (scale of corrections $\pm$10\%), while $S2_{c}$ is the $S2$ signals size corrected to equalize the response to that at the surface (scale of correction from 0 to 50\%). 
$g_{1} = 0.117 \pm 0.003$~phd/photon and $g_{2} = 12.1 \pm 0.8$~phd/electron are the gain factors~\cite{jr:LUXresol}, defined by the expectation values $\left\langle S1 \right\rangle = g_{1} n_{\gamma}$ and $\left\langle S2 \right\rangle = g_{2} n_{e}$, where $n_{\gamma}$ and $n_{e}$ represent the initial number of photons and electrons produced by the interaction; $\epsilon = 49\% \pm 3\%$~\cite{jr:LUXresol} is the efficiency for extracting electrons from the liquid to the gas; and $W = (13.7 \pm 0.2)$~eV~\cite{jr:LUXresol} is the work function for the production of either a photon or an electron.

\section{Analysis}

\subsection{Profile Likelihood Ratio analysis}

\begin{figure}[h!]
\centering
\begin{minipage}[l]{0.5\textwidth}
    \includegraphics[width=0.9\textwidth]{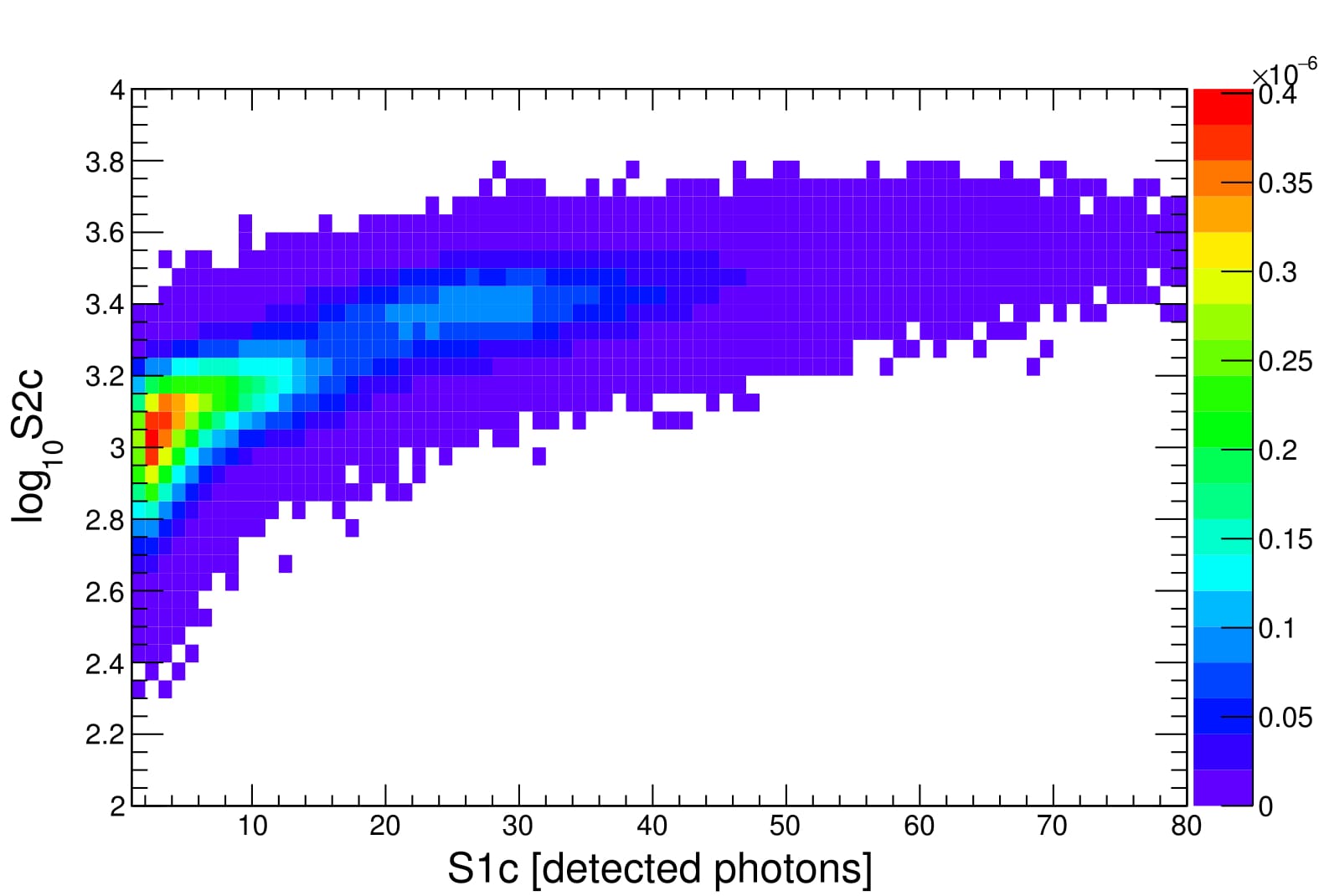}
    \includegraphics[width=0.9\textwidth]{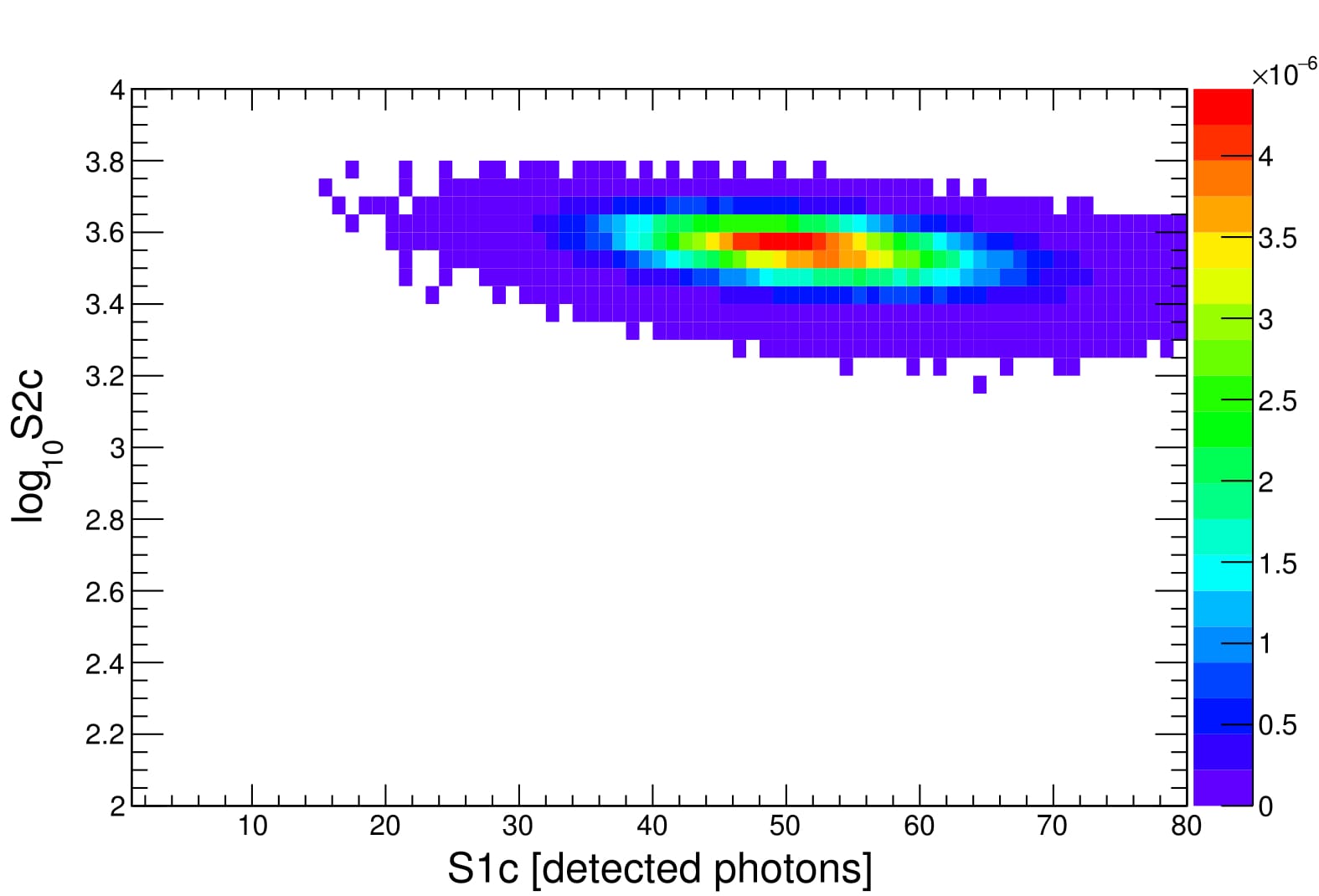}
  \captionof{figure}{Signal models projected on the two-dimensional space of $\log_{10}S2_{c}$ as a function of $S1_{c}$, for massless solar axions (top) and 10 keV/$c^{2} $ mass galactic ALPs (bottom).}
   \label{fig:signalModels}
\end{minipage}
\end{figure}

A two-sided profile likelihood ratio (PLR) analysis \cite{jr:PLRformulae} has been performed to test the 
signal models against the LUX 2013 data. 
The approach used is consistent with that applied to the LUX standard WIMPs search \cite{jr:reanaLUX}, in which the PLR is based on the 
simultaneous separation of the signal and the background distributions in the four physical observables: $r$, $z$, $S1_{c}$, and $\log_{10}(S2_{c})$. 
Conversion of theoretical axion and ALP energy spectra to probability density functions for each of the physical observables has been performed with 
NEST \cite{jr:NEST1, jr:NEST2, jr:yields}, taking into account the detector response and the efficiency.
The models of the signal for the solar axions, and for an example 10~keV/$c^{2} $ mass galactic ALP, are shown in Fig.~\ref{fig:signalModels}, projected on the two-dimensional space of $\log_{10}S2_{c}$ as a function of $S1_{c}$.

Systematic uncertainties in background rates are treated as nuisance parameters in the PLR. Table~\ref{tabular:nuisance} summarizes the contributions from the  background sources, listing the number of events expected in the total exposure and the best fit value returned by the PLR (in the solar axion search). The constraints are Gaussian distributions, with means and standard deviations indicated.

\begin{center}  
\begin{tabular}{cccc}
\hline
\hline
Parameter & Constraint & \parbox[c][1cm]{3cm}{\centering Fit value\\(solar axions)}\\
\hline
Low-$z$-origin $\gamma$ counts & 161 $\pm$ 69 & 157 $\pm$ 17 \\
Other $\gamma$ counts & 223 $\pm$ 96 & 175 $\pm$ 18 \\ 
$\beta$ counts & 67 $\pm$ 27 & 113 $\pm$ 18 \\
$^{127}$Xe counts & 39 $\pm$ 12 & 42 $\pm$ 8 \\
\hline
\end{tabular}
\captionof{table}{Nuisance parameters in the best fit to the 2013 LUX data for solar axions.
Constraints are Gaussian with means and standard deviations indicated. Events counts are after analysis cuts and thresholds, as described in Ref.~\cite{jr:reanaLUX}.}
\label{tabular:nuisance}
\end{center}

The PLR analysis extracts a 90\% C.L. upper limit on the number of signal events: if the local $p$ value is below 10\%, the signal hypothesis is excluded at 90\% C.L.
The limit on the number of signal events is then converted to a limit on the coupling constant between axion/ALP and electrons, $g_{Ae}$.

\subsection{The Look Elsewhere Effect}

\begin{figure}[htb]  
\begin{center}
\includegraphics[scale=0.15]{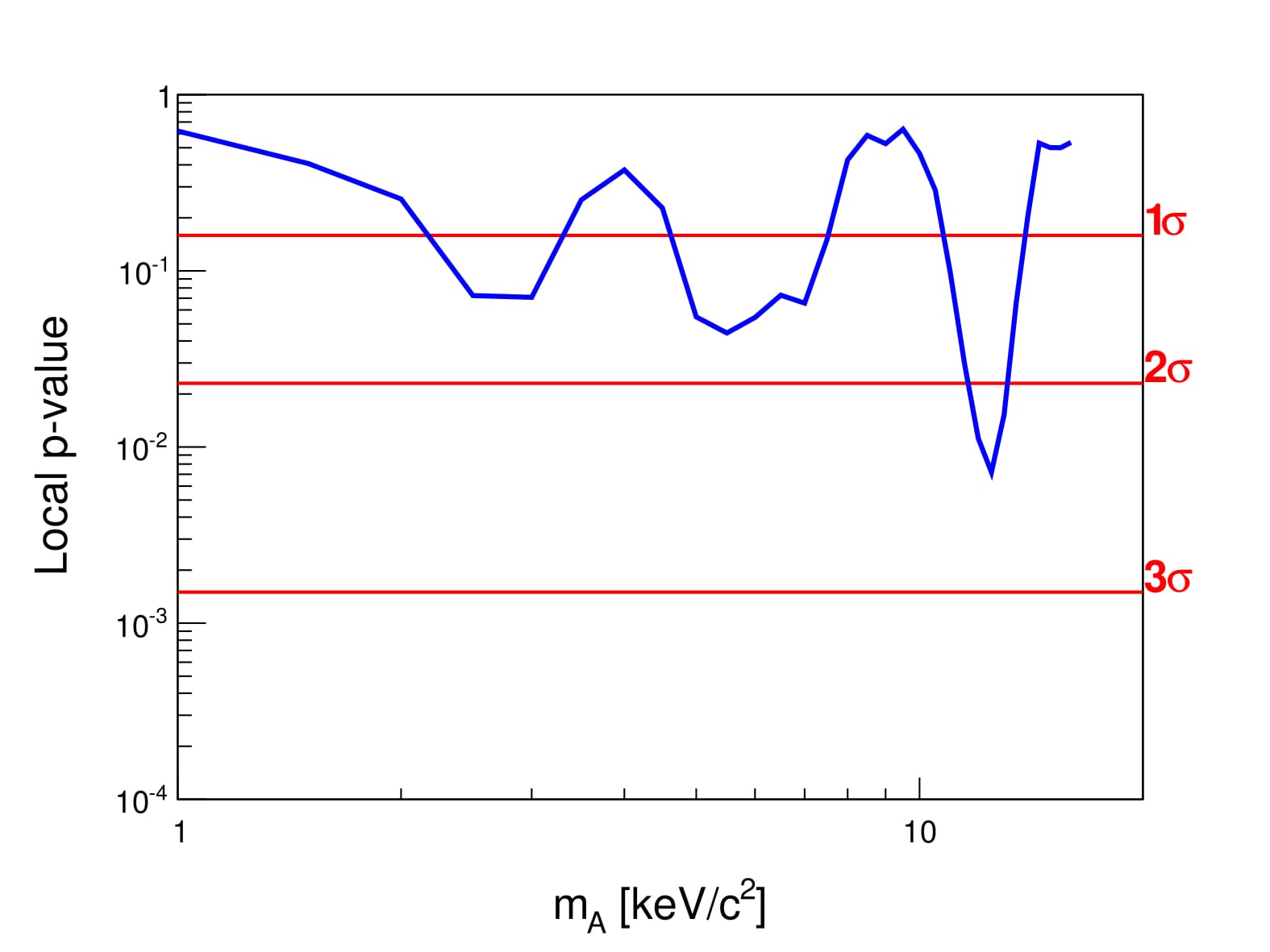}
\caption{Local $p$ value as a function of the ALP mass. The minimum is reached at 12.5 keV/$c^{2}$, where the local $p$ value is 7.2$\times$10$^{-3}$, corresponding to a 2.4$\sigma$ local deviation.}
\label{fig:LocalPvalueALP} 
\end{center}
\end{figure}

The ALP study is conducted by searching for a specific feature over a range of masses. The local significance of observing such a feature at 
one particular mass must be moderated by the number of trials undertaken, in order to calculate a global significance~\cite{jr:LEEgross}.
In Fig.~\ref{fig:LocalPvalueALP}, the local $p$ value, {\it i.e.}, the probability of such an excess if there is no ALP signal at that mass, is plotted as a function of the ALP mass, highlighting the correspondence with the number of standard deviations ($\sigma$) away from the null hypothesis. At 12.5 keV/$c^{2}$ a local $p$ value of 7.2$\times$10$^{-3}$ corresponds to a 2.4$\sigma$ deviation. Following the procedure outlined in Ref.~\cite{jr:HiggsCMS} (where it was applied to searches
for the Higgs boson), a boost factor has been calculated that evaluates the likelihood of finding a deviation for a number of searches as compared to
the significance that would apply to a search performed only once. 
Consequently, the global $p$ value is evaluated as 5.2$\times$10$^{-2}$ at 12.5~keV/$c^{2}$, corresponding to a 1.6~$\sigma$ rejection of the 
null hypothesis.

\section{Results}

\begin{figure}[h]  
\begin{center}
\includegraphics[scale=0.15]{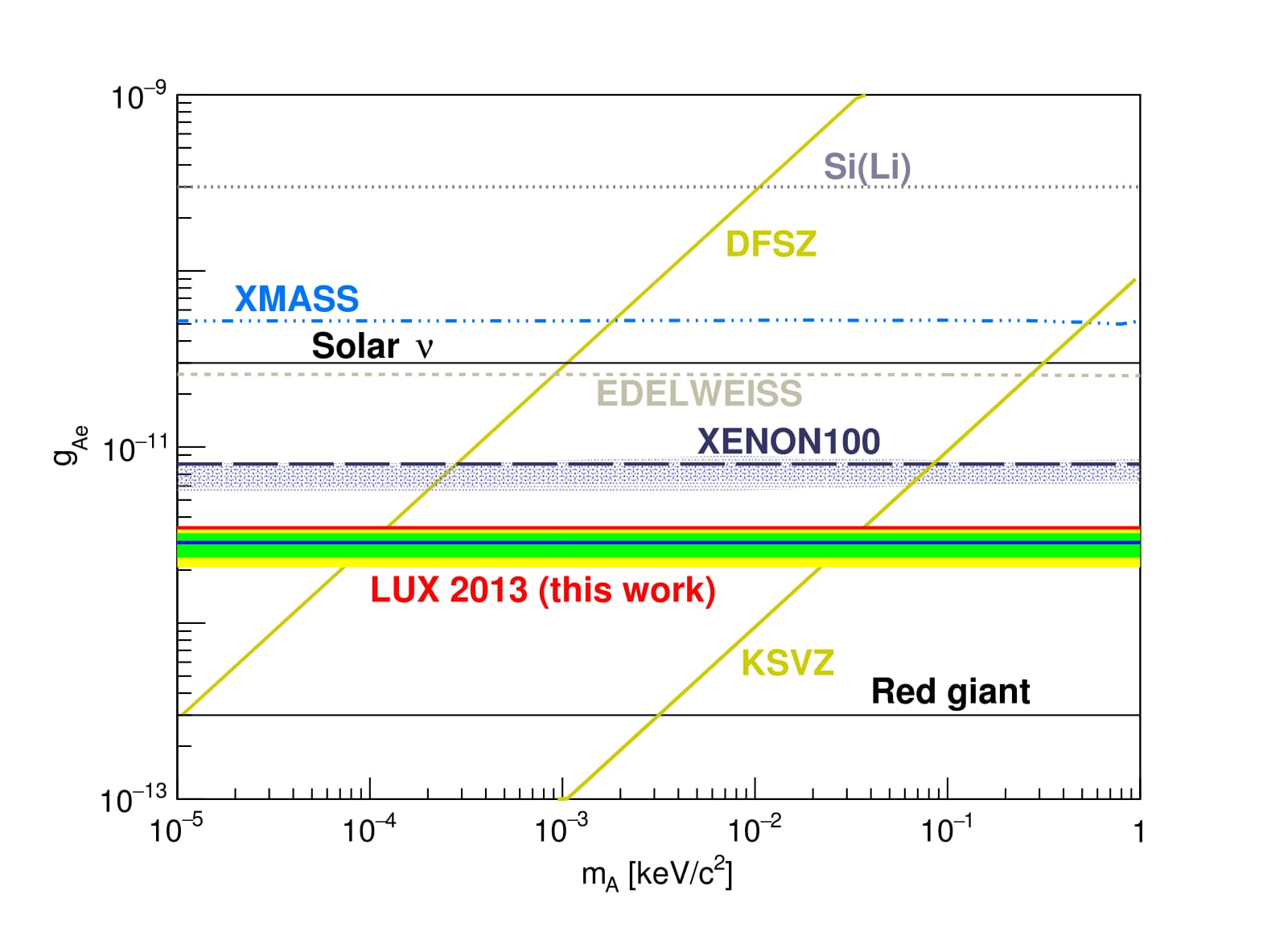}
\caption{Red curve: LUX 2013 data 90\% C.L. limit on the coupling between solar axions and electrons. Blue curve: 90\% C.L. sensitivity, $\pm$ 1~$\sigma$ (green band), and  $\pm$ 2~$\sigma$ (yellow band).}
\label{fig:LimitSolar} 
\end{center}
\end{figure}

The 90\% C.L. upper limit on the coupling $g_{Ae}$ between solar axions and electrons is shown in Fig.~\ref{fig:LimitSolar}, along with the limits set by the previous experiments~\cite{jr:XENON100, jr:EDELWEISS, jr:XMASS, jr:SolarNu}, the astrophysical limit set via the Red Giant cooling process~\cite{jr:RedGiants} and the theoretical models describing QCD axions~\cite{jr:DFSZ, jr:KSVZ1, jr:KSVZ2}. The 2013 LUX data set excludes a coupling larger than 3.5$\times$10$^{-12}$ at 90\%~C.L, the most stringent such limit so far reported.
Assuming the Dine-Fischler-Srednicki-Zhitnitsky model, which postulates the axion as the phase of a new electroweak singlet scalar field coupling to a new heavy quark, 
the upper limit in coupling corresponds to an upper limit on axion mass of 0.12 eV/$c^{2}$, while for the Kim-Shifman-Vainshtein-Zhakharov description, which assumes the axion interacting with two Higgs doublets rather than quarks or leptons, masses above 36.6 eV/$c^{2}$ are excluded. 

\begin{figure}[h]  
\begin{center}
\includegraphics[scale=0.15]{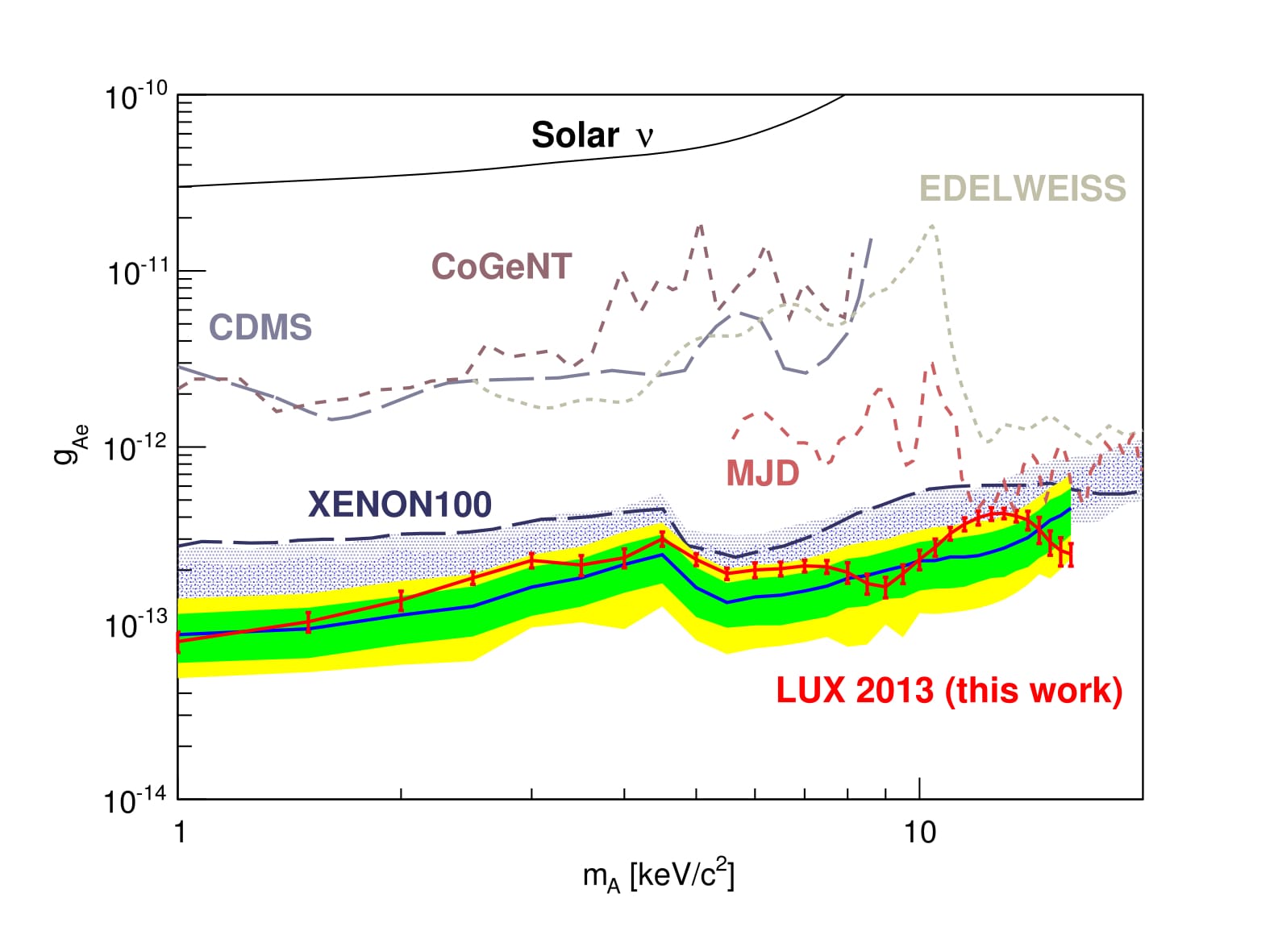}
\caption{Red curve: LUX 2013 data 90\% C.L. limit on the coupling between galactic axion-like particles and electrons. Blue curve: 90\% C.L. sensitivity, $\pm$ 1~$\sigma$ (green band), and  $\pm$ 2~$\sigma$ (yellow band).}
\label{fig:LimitALP} 
\end{center}
\end{figure}

In the galactic ALP study, a scan over masses has been performed, within the range of 1--16~keV/$c^{2}$, limited by the range over which 
precise knowledge of light and charge yield is determined through tritiated methane calibration data~\cite{jr:LUXeffER}. Assuming that ALPs constitute all of the galactic dark matter, the 90\% C.L. upper limit on the coupling between ALPs and electrons is shown in Fig.~\ref{fig:LimitALP} as a function of the mass, together with the results set by other experiments~\cite{jr:MJD, jr:XENON100, jr:CDMS, jr:CoGeNT, jr:EDELWEISS, jr:SolarNu}. 
Again, this is the most stringent such limit so far reported in this mass range.

\section{Summary}

We have presented the results of the first axion and ALP searches with the LUX experiment. Under the assumption of an axio-electric effect interaction in xenon, we test the coupling constant between axions and ALPs with electrons, $g_{Ae}$, using data collected in 2013, for a total exposure of 95 live days $\times$ 118 kg.
Using a profile likelihood ratio statistical analysis, for solar axions we exclude $g_{Ae}$ larger than 3.5$\times$10$^{-12}$ (90\% C.L.) and axion masses larger than 0.12 or 36.6 eV/$c^{2}$ under the assumption of the Dine-Fischler-Srednicki-Zhitnitsky or Kim-Shifman-Vainshtein-Zhakharov theoretical models, respectively.  
For axion-like particles, a scan over masses within the range 1--16~keV/$c^{2}$ excludes discovery of a signal with a global significance at a level of 1.6~$\sigma$, and constrains values of the coupling to be no larger than 4.2$\times$10$^{-13}$, across the full range.

\section{Acknowledgements}

This work was partially supported by the U.S. Department of Energy under Awards No. DE-AC02-05CH11231, DE-AC05-06OR23100, DE-AC52-07NA27344, DE-FG01-91ER40618, DE-FG02-08ER41549, DE-FG02-11ER41738, DE-FG02-91ER40674, DE-FG02-91ER40688, DE-FG02-95ER40917, DE-NA0000979, DE-SC0006605, DE- SC0010010, and DE-SC0015535, the U.S. National Science Foundation under Grants No. PHY-0750671, PHY-0801536,  PHY-1003660,  PHY-1004661, PHY-1102470,  PHY-1312561,  PHY-1347449,  PHY-1505868, and  PHY-1636738, the Research Corporation Grant No. RA0350, the Center for Ultra-low Background Experiments in the Dakotas, and the South Dakota School of Mines and Technology. LIP-Coimbra acknowledges funding from Funda\c c\~ao para a Ci\^encia e a Tecnologia through the Project-Grant No. PTDC/FIS-NUC/1525/2014. Imperial College and Brown University thank the UK Royal Society for travel funds under the International Exchange Scheme (Grant No. IE120804). The UK groups acknowledge institutional support from Imperial College London, University College London and Edinburgh University, and from the Science \& Technology Facilities Council for Ph.D. studentships Grants No. ST/K502042/1 (A. B.), ST/ K502406/1 (S. S.), and ST/M503538/1 (K. Y.). The University of Edinburgh is a charitable body registered in Scotland, with Registration No. SC005336. 
We gratefully acknowledge the logistical and technical support and the access to laboratory infrastructure provided to us by SURF and its personnel at Lead, South Dakota. SURF was developed by the South Dakota Science and Technology Authority, with an important philanthropic donation from T. Denny Sanford, and is operated by Lawrence Berkeley National Laboratory for the Department of Energy, Office of High Energy Physics.

\bibliographystyle{unsrt}     
\bibliography{AxionLiterature}

\end{document}